\documentclass[twocolumn]{openjournal}

\usepackage{xcolor}
\usepackage{textgreek}
 \usepackage{csquotes}
\usepackage[utf8]{inputenc}
\usepackage[english]{babel}
\usepackage{hyperref}
\hypersetup{
    unicode, 
    colorlinks=true,
    linkcolor=linkcolor,
    citecolor=linkcolor,
    filecolor=linkcolor,
    urlcolor=linkcolor,
}
\usepackage{color,colortbl}
\definecolor{linkcolor}{rgb}{0.0,0.3,0.5}
\usepackage{tensind}
\tensordelimiter{?}
\DeclareGraphicsExtensions{.bmp,.pdf,.jpg,.pdf}
\usepackage{verbatim}
\usepackage[normalem]{ulem}
\usepackage{orcidlink}
\usepackage{soul}
\newcommand{\msun}{{\rm M}_{\odot}}
\newcommand{\mstar}{{M}_{\star}}
\newcommand{\rperi}{R_{\rm{peri}}}
\newcommand{\tperi}{t_{\rm{peri}}}
\newcommand{\rnow}{R_{\rm{now}}}
\newcommand{\tquench}{t_{\rm{quench}}}
\newcommand{\vmax}{\Delta V_{\mathrm{max}}}

\graphicspath{ {./figs/} }

\begin{document}
\title{Linking Orbital History to the Quenching of Isolated Dwarf Galaxies}

\shortauthors{Um, Baxter \& Nadler}
\author{Sophia Um$^{*,1,2}$\orcidlink{0009-0003-5057-5371}}
\altaffiliation{$^*$\href{mailto:sum2770@sdsu.edu}{\textcolor{blue}{sum2770@sdsu.edu}}}

\author{Devontae C. Baxter$^{\dagger,1,3}$\orcidlink{0000-0002-8209-2783}}
\altaffiliation{$^{\dagger}$NSF Astronomy and Astrophysics Postdoctoral Fellow}

\author{Ethan O. Nadler$^1$\orcidlink{0000-0002-1182-3825}}
\affiliation{$^1$Department of Astronomy \& Astrophysics,
University of California, San Diego, 9500 Gilman Dr, La Jolla, CA 92093, USA}
\affiliation{$^2$Department of Astronomy, San Diego State University, 5500 Campanile Dr, San Diego, CA 92182, USA}
\affiliation{$^3$Department of Astrophysical Sciences, Princeton University, 4 Ivy Lane, Princeton, NJ 08544, USA}

\begin{abstract}
    Recent discoveries of isolated dwarf galaxies ($\mstar \sim 10^{7-9}~\msun$) with no signs of ongoing star formation have challenged the prevailing notion that these solitary systems are exclusively star-forming. Investigating the origins of isolated quenched dwarfs provides important clues into the dominant processes driving galaxy quenching as a function of environment and stellar mass. In this study, we identify central dwarf galaxies in the Illustris-TNG50 cosmological simulation with $\mstar \sim 10^{7-9}~\msun$, evaluate their star formation and orbital histories, and investigate how their dark matter halo properties correlate with quenching and interactions with massive neighbors. We find that $\rperi$, defined as the closest separation ever attained between a central dwarf and a galaxy more massive than $\mstar > 10^{10}~\msun$, is a reliable metric for inferring the star formation history of central dwarfs in TNG50. We also show that isolated quenched dwarf galaxies separate into backsplash and non-backsplash subpopulations as a function of $\rperi$, and that non-backsplash galaxies are consistent with being quenched by cosmic web stripping or internal feedback mechanisms. Furthermore, we find that $\rperi$ correlates strongly with dark matter halo mass and star formation proxies based on halo maximum circular velocity, which are commonly used in empirical galaxy--halo connection models. These predictions can be tested with data from upcoming deep and wide surveys (e.g., Rubin LSST and Roman) that are anticipated to significantly increase the cosmic census of dwarf galaxies. 
\end{abstract}


\maketitle

\section{Introduction}
\label{sec:intro}

Dwarf galaxies, typically defined as dark-matter dominated systems with $\mstar \leq 10^9~\msun$, are the most abundant galaxies in the Universe \citep{Binggeli90} and crucial building blocks of more massive galaxies due to the hierarchical nature of galaxy formation \citep{WhiteFrenk91}. Relative to their more massive counterparts, dwarf galaxies exhibit exceptionally high dark matter fractions, a quality that makes them unique laboratories for testing predictions from $\Lambda$CDM and alternative theories of dark matter in the smallest luminous halos~\citep{2017ARA&A..55..343B,2022NatAs...6..897S}. Additionally, the relatively shallow gravitational potentials of dwarf galaxies make them exceptional testbeds for constraining the impact of baryonic feedback and environment on driving galaxy evolution.

A key takeaway gleaned from previous deep- and wide-field extragalactic surveys is that the star formation activity of dwarf galaxies is strongly correlated with their local environment. Relative to their stellar mass-matched counterparts in low-density environments, dwarf galaxies that are satellites of a more massive host galaxy are found to have higher instances of suppressed (or \textquote{quenched}) star formation activity \citep[e.g.,][]{Wetzel13, Phillips14, SlaterBell14, Baxter21, Meng23, Greene23, Geha24} and exhausted gas reservoirs \citep[e.g.,][]{Geha06, Putman21}. Data from these surveys also support the prevailing notion that \textquote{isolated} dwarf galaxies---i.e., systems beyond the sphere of influence (e.g., the virial or tidal radius) of another nearby galaxy of equal or greater stellar mass---are almost exclusively star forming down to stellar masses of $\sim 10^7~\rm{M}_{\odot}$ \citep{Haines08, Weisz11, Geha12}. For example, the \cite{Geha12} analysis of Sloan Digital Sky Survey (SDSS) data, and more recently, the \cite{KadoFong25} analysis of Satellites Around Galactic Analogs background (SAGAbg) data, found that the quenched fraction of non-satellite galaxies declines monotonically with decreasing stellar mass, reaching nearly zero in the stellar mass range $10^{8.5} - 10^9~\rm{M}_{\odot}$. A naïve extrapolation of these findings would suggest that lower-mass isolated dwarfs are almost exclusively star-forming. However, observations over the last decade have challenged this notion through the discovery of numerous isolated quenched dwarf galaxies with stellar masses down to $\mstar\sim10^{7}~\msun$ \citep[e.g.,][]{MartinezDelgado16, Makarov17, Polzin21, Sand22, Casey23, Li24, McQuinn24, Sand24, Bennet25, Bidaran25, Paudel25, Luber25, Fielder25, Kaviraj25, Shapiro25, Carlsten26, Hai26}.   

These findings motivate questions regarding the stellar mass limit at which the quenched fraction reaches a minimum before increasing again, and the dominant physical mechanism responsible for suppressing star formation in isolated dwarf galaxies. Some studies suggest that baryonic feedback processes (e.g., supernova feedback) are potentially responsible for quenching isolated dwarf galaxies \citep[e.g.,][]{Fillingham18, SandovalAscencio25, Lazar26}. This quenching pathway is expected to be most effective in the regime of bright and classical dwarf galaxies, where stellar masses are high enough to enable strong feedback and the underlying gravitational potential wells are shallow (e.g., \citealt{Hayashi26}). Another quenching pathway effective in the dwarf galaxy stellar mass regime is the UV photoionizing background produced during the Epoch of Reionization. In this scenario, the radiation field produced during reionization quenches low-mass galaxies by heating and ejecting their star-forming gas, though this mechanism is expected to be most effective for ultra-faint dwarf galaxies with $\mstar < 10^{5}~\msun$ \citep[e.g.,][]{Efstathiou92, Thoul96, Barkana99, Tollerud18, Fillingham19,  Wheeler19, RodriguezWimberly19}. Nevertheless, there is debate as to whether reionization is sufficient to quench more massive dwarfs ($\mstar \sim 10^{7-9}~\msun$) found in the field \citep{DiCintio17, Fillingham18, Benavides21, Chan18}. Other quenching pathways typically require gravitational or hydrodynamical interactions with a more massive galaxy, as is the case for tidal-stripping \citep{Merritt83, Moore99, Gnedin03} and ram-pressure stripping \citep{GG72, Abadi99, Emerick16}, respectively. While dwarf--dwarf gravitational interactions occur more frequently than interactions with massive galaxies, observational evidence from \citet{Stierwalt15} suggests these interactions do not significantly contribute to the quenching of central dwarf galaxies (but also see \citealt{KadoFong24}). Thus, with the exception of internal feedback mechanisms, the aforementioned quenching pathways are unlikely to fully explain recent observations of isolated quenched dwarf galaxies with $\mstar \sim 10^{7-9}~\msun$ \citep{Roman19, Prole21, SandovalAscencio25}.  

Cosmological hydrodynamical simulations and semi-analytic models have served as indispensable tools for interpreting the physical drivers behind observed quenched fractions \cite[e.g.,][]{Wheeler14, Fillingham15, Akins21, Baxter23, Baxter25} and making testable predictions for the abundance of quenched galaxies as a function of redshift, stellar mass, and environment \citep[e.g.,][]{Behroozi2019, Donnari19, DeLucia24, Xie24, Doven26}. While modern models generally reproduce the observed quenched fractions for massive central and satellite galaxies at $z=0$ \citep[e.g.,][]{Xie20, Donnari21},  predictions for the quenched fraction of dwarf galaxies and their underlying quenching mechanisms often vary across hydrodynamic simulation suites and between simulations and empirical models. For instance, \citet{Mercado26} recently demonstrated that the quenched fraction as a function of stellar mass for dwarf galaxies ($\mstar = 10^{7-9}~\msun$) around Milky Way-like hosts in the FIREbox~\citep{FIREbox}, FIRE-2~\citep{FIRE2}, and TNG50~\citep{TNG50_I, TNG50_II} cosmological simulations are broadly consistent with observations from the Satellites Around Galactic Analogs \citep[SAGA,][]{Geha17} and  Exploration of Local VolumE Satellite \citep[ELVES,][]{Carlsten2022} surveys. However, these simulations predict vastly different quenched fractions of dwarf satellites as function of host-centric radius, highlighting that the environmental dependence of dwarf galaxy quenching is relatively sensitive to the underlying physical model. Similarly, \cite{Dickey21} compared the EAGLE~\citep{EAGLE} and SIMBA~\citep{SIMBA} cosmological simulations to estimate the quenched fraction in isolated dwarfs with $\mstar =10^{8-9}~\msun$. They found that EAGLE predicts a quenched fraction that rises from $\sim 0.1$ at $\mstar=10^{9.5}~\msun$ to $\sim 0.35$ at $\mstar=10^{8}~ \msun$, while SIMBA predicts that the quenched fraction never rises above 0.1 for $M_*<10^9~\msun$, suggesting that dense environments are necessary for quenching in dwarfs. Predictions have also been made using empirical models; for example, \cite{Wang2024} used a version of the empirical UniverseMachine model~\citep{Behroozi2019} calibrated to SAGA satellite population data to predict the quenched fraction of isolated dwarf centrals down to $M_*\approx 10^{6.5}~\msun$. They found that the quenched fraction begins to rise steeply to $\sim0.3$ at stellar masses $M_*\leq 10^{7.5} \msun$, and that quenching for isolated dwarfs is driven almost entirely by halo assembly, meaning that quenched galaxies had their star formation halted at early times due to interactions with other galaxies and did not resume star formation thereafter.

Despite their quantitative differences, the results above consistently imply that the environments experienced by dwarf galaxies throughout cosmic time---i.e., their orbital histories throughout the cosmic web and within larger hosts---play a key role in the quenching process. In this context, dwarf galaxies that are isolated at $z=0$ are particularly informative, since they represent a pristine population that potentially never interacted with a more massive galaxy. Thus, comparisons between the properties of extremely isolated dwarfs with their counterparts in denser environments can be used to gain insights into the drivers of galaxy quenching. 

This work will use data from the TNG50 simulation to further investigate the quenching mechanisms and environmental histories of isolated dwarfs as a function of their historical distance to any more massive host galaxy. In particular, we categorize isolated quenched dwarfs into two populations that have been identified in previous studies: backsplash and non-backsplash systems. Backsplash occurs when a galaxy orbits within the virial radius of a more massive host and subsequently escapes (e.g., \citealt{Teyssier12}); pericentric passages during this orbital interaction can tidally strip gas from dwarf galaxies, shutting down their star formation \citep[]{Wetzel13, SlaterBell14, Wetzel15, Simpson18, Putman21}. On the other hand, non-backsplash quenched dwarfs may be quenched by hydrodynamical interactions with cosmic web filaments or internal feedback mechanisms. The former is referred to as cosmic web-stripping and occurs when the star-forming gas of a galaxy is stripped via ram-pressure as it passes through the diffuse gas of the cosmic web \citep{Benitez-Lambay13,Benavides25}. Meanwhile, for dwarfs with high gas-mass fractions, strong gas outflows driven by stellar feedback can prevent further gas cooling and collapse, resulting in internal quenching (e.g., \citealt{Wu26}).

Although these two classes of isolated dwarfs (backsplash and non-backsplash) both contain quenched galaxies, the split between these populations is expected to depend on other physical properties, including mass, velocity, and degree of isolation, potentially resulting in distinguishable observational signatures. Our goal is to use cosmological simulations to identify observables that encode the orbital and star formation histories of isolated dwarfs to more cleanly separate backsplash and non-backsplash systems and constrain their quenching pathways. This work is timely as identifying these observables is important for the interpretation of upcoming data from new facilities, such as the Vera C.\ Rubin Observatory and Nancy Grace Roman Space Telescope, that are anticipated to discover an unprecedented number of dwarf galaxies across a range of environments \citep{Shread26, Tsiane25, MutliPakdil21,Sanderson2026}. Data from these next-generation telescopes will be used to infer the sizes, stellar and halo masses, and redshifts of distant, faint dwarfs, all of which can be measured in simulations and correlated with parameters that capture the orbital and star formation histories of dwarf galaxies.

This paper is organized as follows. In \S\ref{sec:sims}, we describe the selection criteria used to construct our sample of central dwarf galaxies from the Illustris-TNG50 simulation, explain our classification scheme for backsplash galaxies, and define the parameters $\rperi$, $\rnow$, and $\vmax$ that are used throughout our analysis. In \S\ref{sec:environments}, we present our results showing the correlations between $\rperi$ and measures of present-day environment and star formation activity. In \S\ref{sec:physical}, we show how $\rperi$ and $\vmax$ strongly correlate with the baryonic and dark matter properties of dwarf galaxies in TNG50. In \S\ref{sec:discussion}, we contextualize our results with findings derived from other simulation studies as well as recent observational surveys of dwarf galaxies. Lastly, in \S\ref{sec:conclusion}, we summarize our results and present our main conclusions.

\section{Simulations and Methods}
\label{sec:sims}

\begin{center}
    \begin{figure}[!t] 
        \centering
        \hspace*{-0.2in}
    	\includegraphics[trim={0cm 0cm 0cm 1.5cm},scale=0.4]{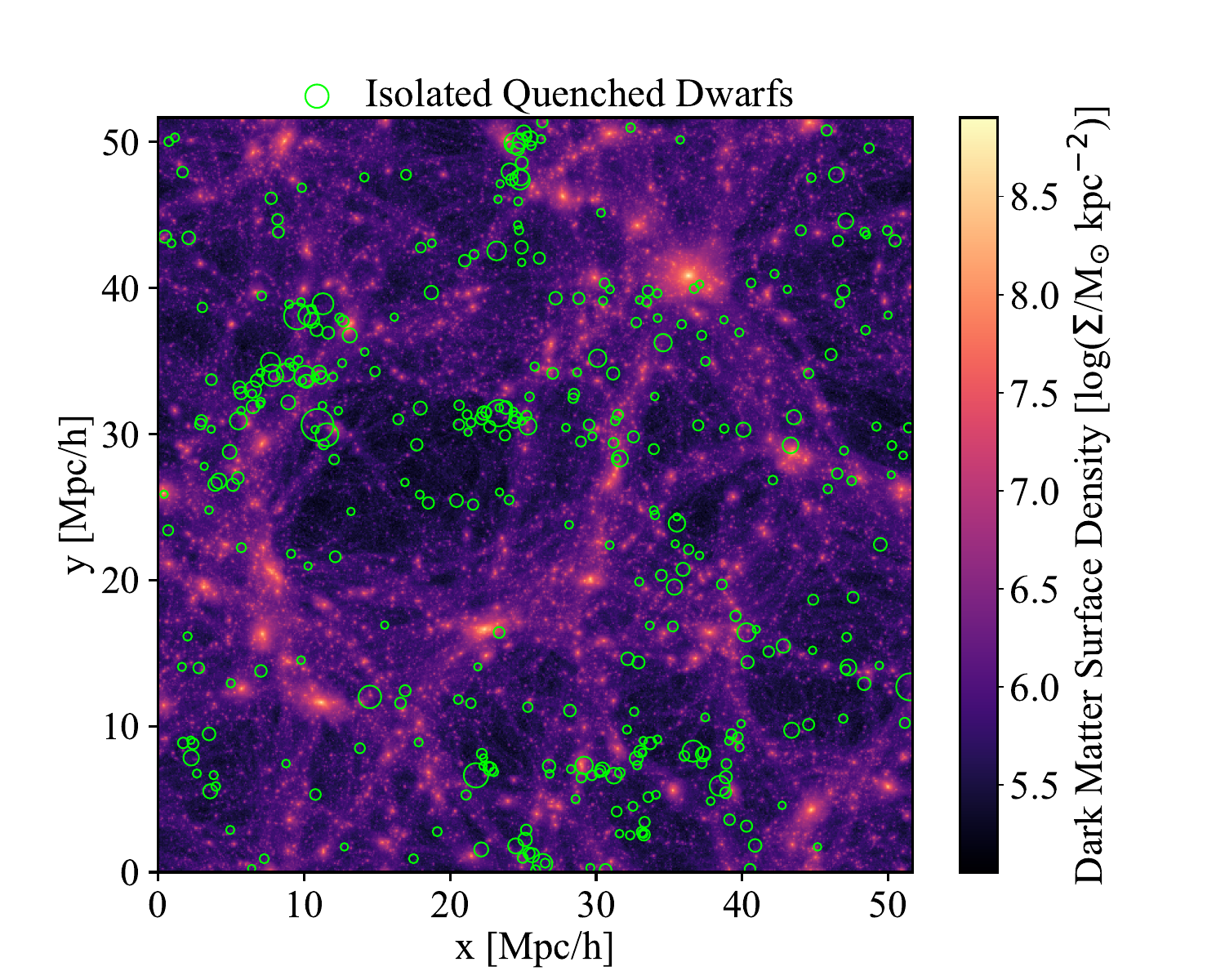}
    	\caption{Projected dark matter surface density in the TNG50 box. Green circles identify the X-Y projected locations of all isolated quenched dwarf galaxies we consider (i.e., systems that are not satellites and have $M_*>10^7~M_{\rm{\odot}}$ and sSFR$<10^{-11}~\mathrm{yr}^{-1}$). The size of the green circles is scaled based on $\log(M_*)$. These systems are preferentially found in voids and along filaments, forming a continuum with backsplash and satellite galaxies.}
        \label{fig:projection}\vspace{0.3cm}
    \end{figure}
\end{center}

We analyze \textit{The Next Generation Illustris Project}\footnote{\href{https://www.tng-project.org}{https://www.tng-project.org}} \citep[IllustrisTNG,][]{Nelson18, Naiman18, Springel18, Pillepich18, Marinacci18}, a suite of magnetohydrodynamical cosmological simulations implemented with the moving-mesh \texttt{AREPO} code \citet{Springel10}. The simulated galaxy population explored in this analysis is selected from the TNG50-1 run (hereafter TNG50), which is the highest resolution iteration from the IllustrisTNG suite with $2\times2160^{3}$ particles (dark matter and baryons) within a periodic cubic box of side length $51.7$ cMpc. At $z=0$, the stars and dark matter have gravitational softening lengths of $0.74~\mathrm{ckpc}$; the gas is adaptively softened with a minimum of $0.18~\mathrm{ckpc}$. This yields an average dark matter and baryonic particle mass resolution of $m_{\rm{DM}}=4.5\times10^{5}~\msun$ and $m_{\rm{baryon}}=8.5 \times 10^{4}~\msun$, respectively. At this resolution, the abundance of dwarf galaxies with stellar masses $M_*\gtrsim10^7~\msun$ is well resolved, and the quenched fraction is consistent with the next-highest-resolution TNG50 run at the $\approx 10\%$ level over the stellar mass range of interest~\citep{Joshi210112226}. As shown in Figure~\ref{fig:projection}, the TNG50 simulation volume contains a range of systems and environments, including galaxy clusters, galaxies with masses comparable to the Milky Way, and voids. As a result, TNG50 allows us to study how the properties of isolated dwarf galaxies correlate with environment.

We construct our sample of isolated (central) dwarf galaxies by first selecting all objects with stellar masses between $\mstar = 10^{7-9}~\msun$ at $z=0$, which have well-resolved star formation rates. Throughout, we measure stellar masses within twice the stellar half-mass radius. We use the stellar half-mass radius rather than the total stellar mass because this quantity more closely tracks the values found using the half-light radius, which is most commonly measured; this choice does not affect the trends we report. In addition, we track the total dark matter halo mass associated with each dwarf galaxy. We then use the \texttt{SUBFIND} merger trees to classify dwarfs as either centrals or satellites at each snapshot based on whether or not they reside within the virial radius ($R_{\rm{200,crit}}$) of another galaxy.\footnote{$R_{\rm{200,crit}}$ is defined as the radius of a sphere with a mean density that is $200\times$ the critical density of the Universe.} We restrict our analysis to centrals; that is, we do not include systems that are satellites at $z=0$ in our sample, regardless of the mass of the parent system. We further classify these isolated dwarfs into backsplash and non-backsplash systems. Specifically, for this sample, we define backsplash galaxies as those centrals that were marked as satellites at any $0<z<10$ redshifts, while non-backsplash galaxies are those that were never marked as satellites back to redshift $z=10$.

We categorize quenching in isolated dwarf galaxies based on their star-formation as follows: quenched galaxies are defined as those that are not forming enough stars to meet the quenching threshold, or:
\begin{equation}
    \mathrm{sSFR} \equiv \frac{\dot{M}_*}{M_*}<10^{-11}~\mathrm{yr}^{-1},\label{eq:sfr_threshold}
\end{equation} 
where $\dot{M}_*$ is the instantaneous star formation rate at $z=0$. In contrast, star-forming galaxies are defined as systems with $\mathrm{sSFR}>10^{-11}~\mathrm{yr}^{-1}$. The trends we report are not sensitive to the specific value of this threshold. For quenched systems, the quenching time $\tquench$ is the largest lookback time when the galaxy permanently dropped below the quenching threshold defined by Eq.~\ref{eq:sfr_threshold}. The green circles in Figure~\ref{fig:projection} show the locations of isolated quenched dwarfs at $z=0$.

Throughout, we measure $\rperi$, the distance of closest approach to a massive galaxy made by a dwarf at any time, as follows:
\begin{equation}
    R_{\mathrm{peri}} \equiv \min_{z<z_i} r(z),
\end{equation}
where $r(z)$ is the 3D comoving distance from a dwarf galaxy to its nearest most massive neighbor as a function of redshift and $z_i=10$ is chosen to avoid early stages of structure formation where host--satellite relationships can be ambiguous. At each snapshot, each dwarf's nearest massive neighbor is defined as the closest system with $M_*>10^{10}~\msun$ at that redshift. The nearest massive neighbor need not be the same system across snapshots. We measure $\tperi$ as the corresponding lookback time at which $\rperi$ occurred, and we always report $R_{\mathrm{peri}}$ in units of comoving Mpc. Finally, we define $R_{\mathrm{now}}$ as the distance to the nearest massive neighbor at $z=0$. Our result are relatively insensitive to the massive neighbor threshold stellar mass as long as it is sufficiently large ($M_*\gtrsim 10^{9.5}~\msun$).

We go on to compare $\rperi$ to select physical parameters: size, the halo assembly parameter $\Delta V_{\rm{max}}$, stellar and halo mass (and the relation between the two), and quenching time. The size of each dwarf is defined as the radius containing half of its stellar mass. $\Delta V_{\rm{max}}$ is the halo assembly parameter, defined as~\citep{Behroozi2019}:
\begin{equation}
\label{eq:delta_vmax}
    \Delta V_{\rm{max}}(z=0) = \frac{V_{\rm{max}}(z=0)}{V_{\rm{max}}(\rm{max}[\mathit{z}_{\rm{dyn}}, \mathit{z}_{\it{M}_{\rm{peak}}}])}
\end{equation}
where $z_{\rm{dyn}}$ is the redshift one dynamical timescale ago (measured, in our analysis, with respect to $z=0$) and $z_{M_{\rm{peak}}}$ is the redshift when the (sub)halo reached its maximum halo mass. This parameterization tracks halo assembly histories in a continuous way that does not rely on categorizing systems into centrals versus subhaloes and has been shown to correlate with SFR. Thus, $\Delta V_{\mathrm{max}}$ does not rely on explicitly tracking infall times or tidal forces, allowing for direct comparisons between isolated and (former) satellite galaxies.

\section{The Dependence of Dwarf Galaxy Quenching on Pericentric Distance}
\label{sec:environments}

In this section, we investigate how measures of environment correlate with quenching in isolated dwarf galaxies by comparing relations between $\rperi$, $\rnow$, and star-formation activity. 

\subsection{Present-day versus Pericentric Distance}

\begin{center}
    \begin{figure}[!t]
        \centering
        \hspace*{-0.15in}
    	\includegraphics[trim={0cm 0cm 0cm 0cm},scale=0.50]{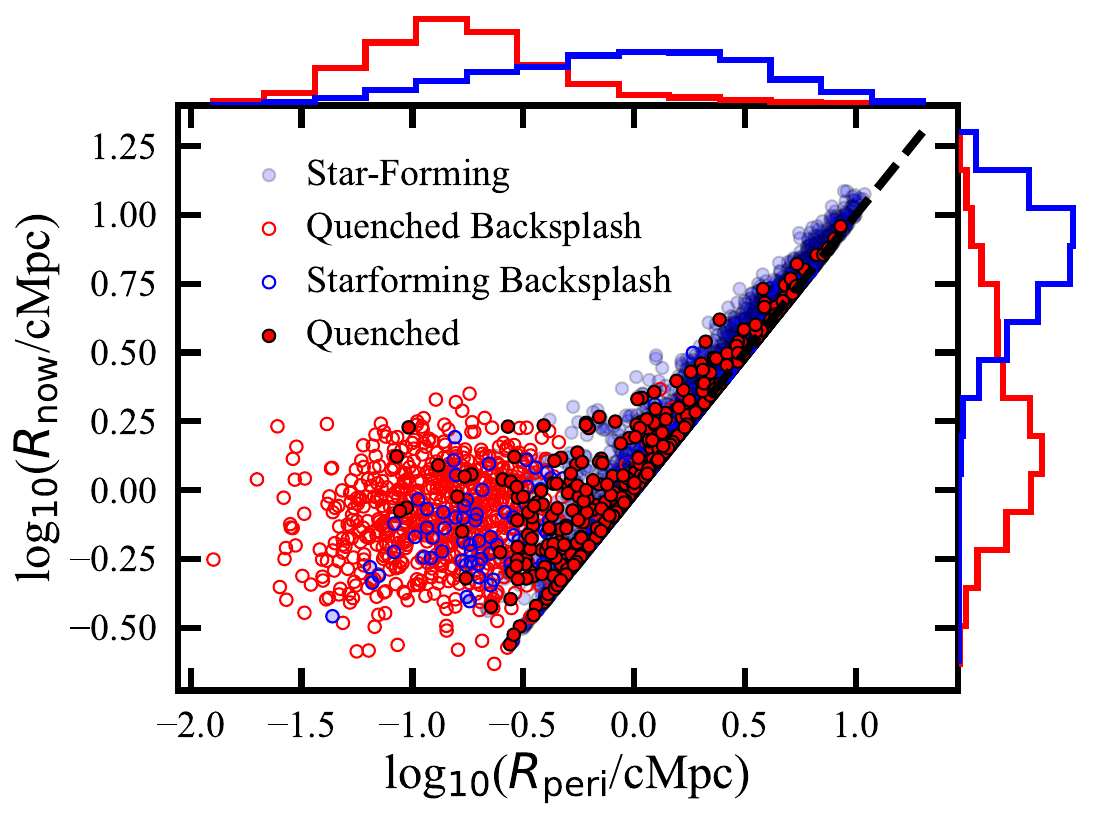}
    	\caption{Current distance of isolated dwarfs to the nearest massive central at $z=0$ ($\rnow$) versus distance of closest approach between a dwarf galaxy and a massive central ($M_*>10^{10}~\msun$) at any time ($\rperi$). The majority of quenched systems exist at $R_{\mathrm{now}}\lesssim 1~\mathrm{cMpc}$, while star-forming populations tend to exist at larger current distances. Galaxies on the one-to-one line are those whose $\rperi$ is equal to their $R_{\mathrm{now}}$, i.e., systems that have not yet fallen into the virial radius of a more massive host. Of all isolated dwarfs, we find 6151 total star-forming galaxies (inclusive of backsplash), 1061 quenched galaxies (backsplash inclusive), 863 backsplash systems (quenched or star-forming), and 6349 non-backsplash systems (quenched or star-forming).}\vspace{0.3cm}
        \label{fig:rnow_rperi}
    \end{figure}
\end{center}

Figure~\ref{fig:rnow_rperi} shows the relationship between $\rperi$ and $\rnow$ for our sample of isolated dwarf galaxies at $z=0$. Dwarfs that satisfy (do not satisfy) our backsplash criteria are shown using unfilled (filled) circles, while the star-forming and quenched subpopulations are colored blue and red, respectively. We find that quenched galaxies exist at all values of $\rperi$ spanned by dwarfs in our sample, and we can identify two distinct evolutionary histories at the extrema of this $\rperi$ distribution. As a reminder, \emph{we exclude satellites from our dwarf sample}, which explains why the lowest values of $\rnow$ we obtain ($\rnow \approx 0.3$ cMpc) are barely larger than the virial radius of a typical Milky Way-mass host halo. Of all isolated dwarfs, we find 6151 total star-forming galaxies (inclusive of backsplash), 1061 quenched galaxies (backsplash inclusive), 863 backsplash systems (quenched or star-forming), and 6349 non-backsplash systems (quenched or star-forming).

Galaxies with low $\rperi$ values ($\rperi \lesssim 0.3$ cMpc) show signs of having experienced a previous close encounter with a massive neighbor followed by ejection from the system in relatively recent history---the larger $\rnow$ values likely reflect the limited time available to travel to greater distances. This interpretation is consistent with our finding that most of these low-$\rperi$ dwarfs are quenched and satisfy our backsplash criterion. The $\rnow$ distribution of these systems also agrees with previous findings from \citet{Bhattacharyya25}, which show that backsplash galaxies in TNG50 are generally $0.5-2$ Mpc away from their host at $z=0$. Most of these backsplash dwarfs had a relatively recent interaction, with an average $\tperi \approx 4.6$ Gyr; we defer further discussion of $\tquench$ and $\tperi$ to \S\ref{quenchingtimes}. In contrast, galaxies with higher $\rperi$ values ($\rperi \gtrsim 0.3$ cMpc) tend to lie closer to the one-to-one relation, suggesting that they have not yet interacted with another massive galaxy and/or may be on their first infall.

We posit that these two subpopulations largely correspond to the two quenching methods identified in the literature (e.g., \citealt{Benavides25}): backsplash galaxies quenched by close encounters with more massive systems (low $\rperi$, low $\rnow$) and cosmic web-stripped galaxies quenched by environmental interactions with large-scale structure such as filaments (high $\rperi$, high $\rnow$). In particular, Figure~\ref{fig:rnow_rperi} implies that nearly all backsplash galaxies with $\rperi\lesssim 100$ ckpc are quenched, consistent with previous results~\citep{Reeves23, Benavides21, Simpson18, Fillingham16, Mayer06}, and that quenching is increasingly common for galaxies further along their infall into a massive host, even if they have not entered its virial radius. We quantify these statements using measurements of quenched fractions in the following subsection.

\begin{center}
    \begin{figure}[!t]
    	\includegraphics[width=0.46\textwidth]{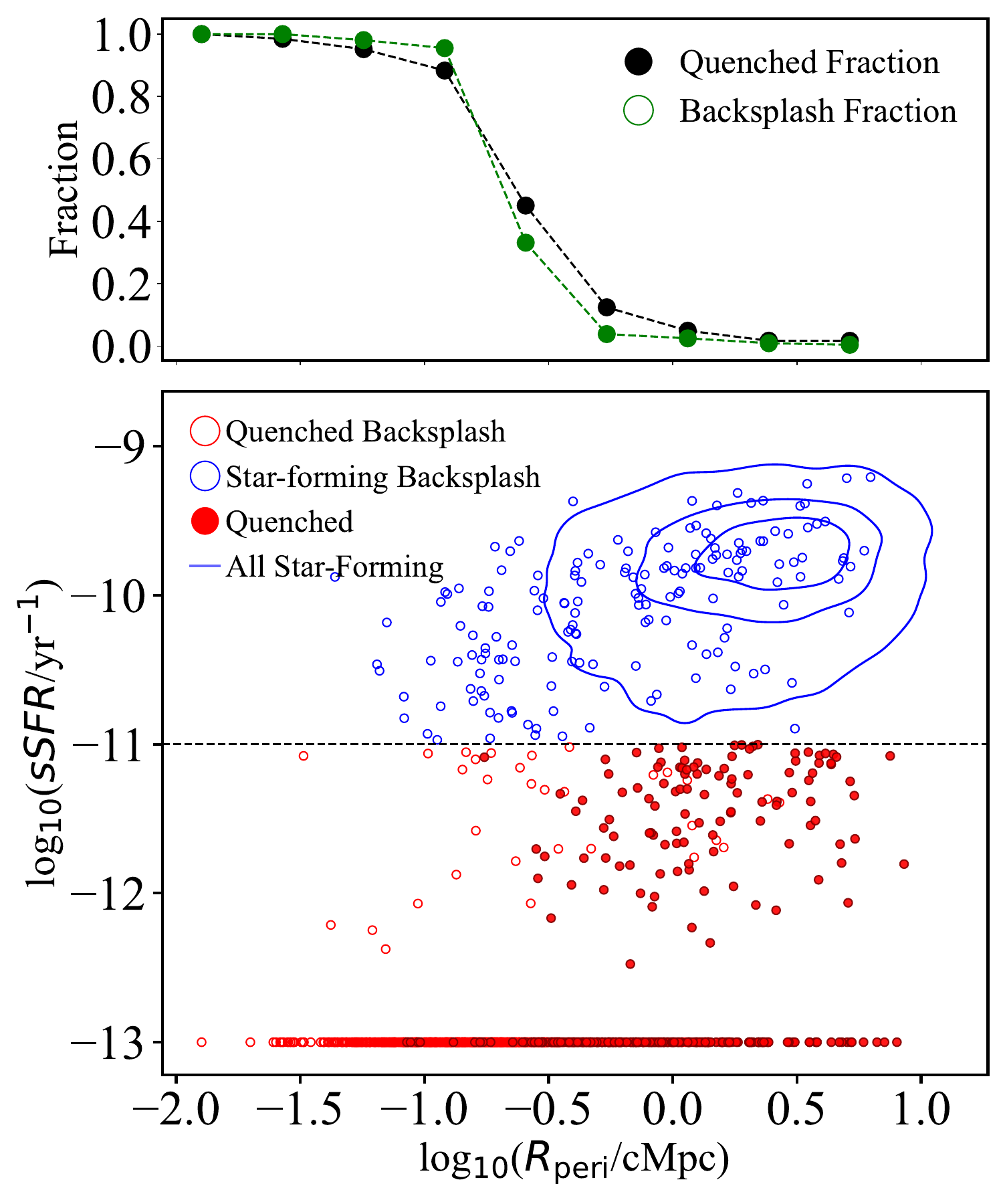}
    	\caption{\emph{Upper}: Quenched (black) and backsplash (green) fractions as a function of $R_{\mathrm{peri}}$. Quenched galaxies are concentrated in the lowest bin ($R_{\mathrm{peri}}\lesssim 0.3~\mathrm{cMpc}$), with only a few quenched galaxies that have $\rperi$ greater than 0.3 cMpc. The backsplash fraction is closely correlated with (but is not identical to) the quenched fraction, and does not exceed the quenched fraction at higher $\rperi$, suggesting a non-backsplash quenching mechanism that does not require close passage to a more massive neighbor. \emph{Lower}: sSFR vs.\ $R_{\mathrm{peri}}$. All galaxies below the black dashed line are defined as quenched (red circles), with non-backsplash (backsplash) systems shown by filled (unfilled) circles. Star-forming backsplash galaxies are shown as open blue circles, and non-backsplash star-forming galaxies are shown by blue isodensity contours with $1$, $2$, and $3\sigma$ levels. As seen in the upper figure, most quenched systems are concentrated at $\rperi$ $<$ 2 cMpc. We also note that star-forming systems, which are only denoted by the isodensity contours, dominate by number at the higher $\rperi$ bins and thus drive the quenched and backsplash fractions down at these bins.}\vspace{0.5cm}
        \label{fig:ssfr_rperi}
    \end{figure}
\end{center}

\begin{center}
    \begin{figure}[!t]
        \centering
    	\includegraphics[scale=0.42]{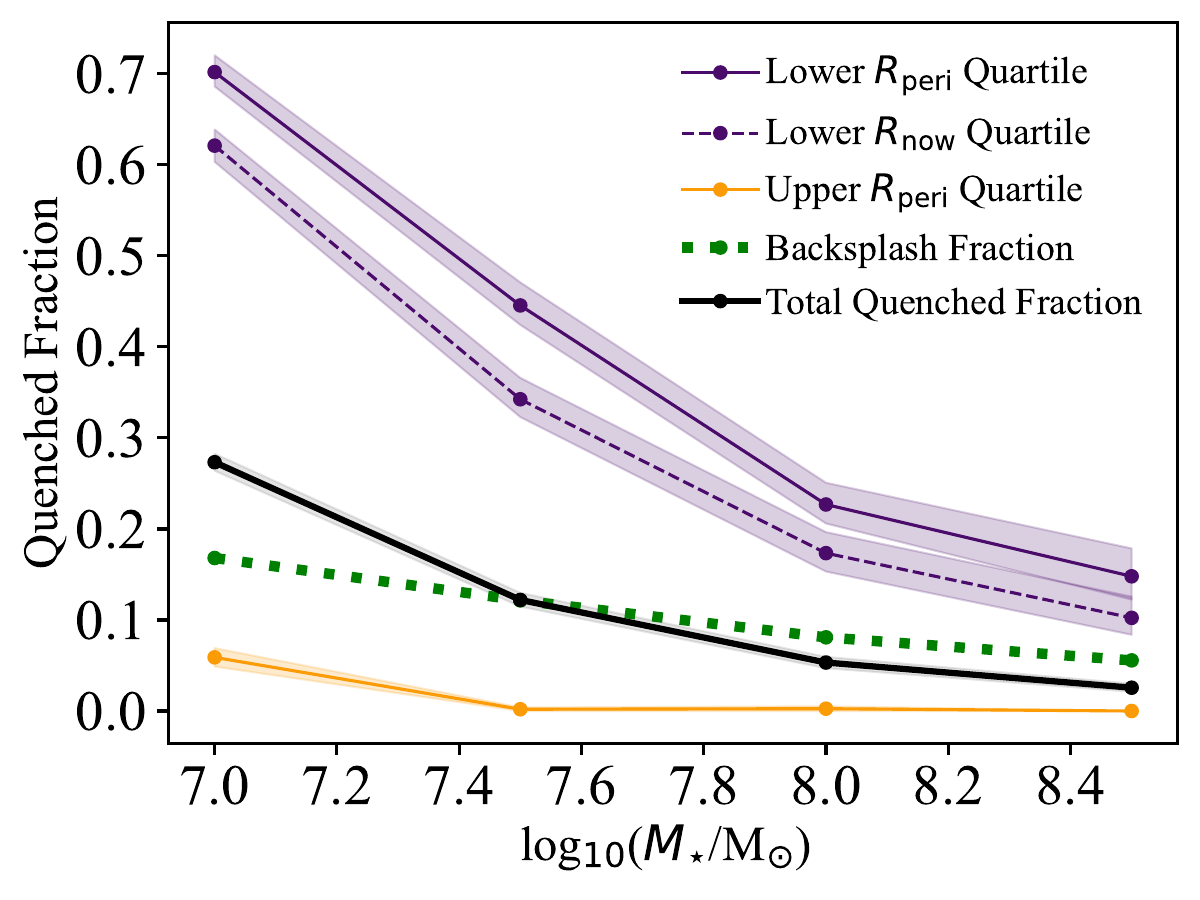}
    	\caption{Quenched fraction vs. stellar mass. Solid (dashed) purple and orange lines represent the lower and upper quartiles of $R_{\mathrm{now}}$ ($R_{\mathrm{peri}}$). The black line represents the entire quenched fraction for the whole isolated dwarf population. The green dotted line represents the backsplash fraction; both the backsplash fraction and quenched fraction rises with decreasing stellar mass. In the upper $R_{\mathrm{now}}$ and $R_{\mathrm{peri}}$ quartiles, $\approx 5\%$ of systems are quenched at the lowest masses. For visual clarity, we exclude the line which corresponds to the upper $\rnow$ quartile (as it is identical to the upper $\rperi$ quartile line). Uncertainties were calculated for each bin using the bootstrap method with 1000 re-samples and finding the 16th (lower) and 84th (upper) percentiles, or $\pm 1 \sigma$, which is distinct from the 25th (lower) and 75th (upper) percentiles used to separate the population by their $\rperi$ values.}
        \label{fig:fqquartiles}\vspace{0.3cm}
    \end{figure}
\end{center}

\subsection{Quenched Fraction versus $\rperi$}

The top panel of Figure~\ref{fig:ssfr_rperi} compares the quenched (black line) and backsplash (green line) fraction of isolated dwarfs as a function of $\rperi$; these fractions track each other closely over the full range of $\rperi$ values in our sample. The quenched fraction rises steeply for $\rperi\lesssim 0.3~\mathrm{cMpc}$; in this range, nearly all quenched galaxies are backsplash systems. For $0.3~\mathrm{cMpc}\lesssim \rperi\lesssim 1~\mathrm{cMpc}$, the quenched fraction is non-zero and is not fully dominated by backsplash systems, implying that a subpopulation of dwarf galaxies quenched without yet experiencing a close interaction with a more massive neighbor. 

According to Figure~\ref{fig:rnow_rperi}, many of these isolated quenched non-backsplash dwarfs are approaching first infall into a more massive system, suggesting that they are quenched by cosmic web-stripping, while a small fraction have large values of $\rperi$ and $\rnow$, suggesting that they are internally quenched. These results are broadly consistent with existing measurements of the quenched fraction, which show that---for extremely isolated dwarf galaxies---the quenched fraction is very small down to $M_*\approx 10^{7.5}~\msun$~\citep{Geha12, KadoFong25}, while measurements in less isolated environments such as the COSMOS field find a larger quenched fraction that rises for $M_*\lesssim 10^8~\msun$~\citep{Lazar26}.  

In the lower panel of Figure~\ref{fig:ssfr_rperi}, we further explore the relationship between star formation activity (parameterized by sSFR) and $\rperi$. Here, the horizontal dashed line represents the threshold adopted in this study to separate star-forming and quenched galaxies (see Eq.~\ref{eq:sfr_threshold}). Likewise, we visually distinguish the star-forming and quenched subpopulations using blue contours and red circles, respectively; we also indicate star-forming backsplash systems as blue open circles. We find that the isolated dwarf galaxies that are quenched have a median $\rperi$ of $\approx0.9$ cMpc, whereas their star-forming counterparts have a median $\rperi$ of $\approx 2.3$ cMpc, which is roughly 2.5 times larger.

Nevertheless, we find an extended tail of isolated quenched dwarf galaxies that spans $\rperi$ values from $\approx2-8$ cMpc, which have thus been far beyond the direct gravitational influence of a more massive system throughout their history. Moreover, we find that the fraction of isolated quenched dwarf galaxies that satisfy the backsplash criterion drops with increasing $\rperi$, an indication that these systems are not quenched due to direct interactions with a significantly more massive neighbor; this population likely contains a mixture of cosmic web-stripped and internally-quenched systems.

\subsection{Quenched Fraction versus Stellar Mass}

In Figure~\ref{fig:fqquartiles}, we plot the quenched fraction as a function of stellar mass for our population of isolated dwarf galaxies. The black line represents the total quenched fraction, while the solid and dashed lines represent the quenched fraction measured only using dwarf galaxies in the lowest (purple, red) and highest (orange) quartiles of $\rperi$  and $\rnow$, respectively. The shaded bands depict uncertainties estimated using bootstrap resampling. The green dotted line represents the backsplash fraction as function of stellar mass. The lowest quartiles ($\leq$ 25th percentile) of $\rperi$ and $\rnow$ for all isolated dwarfs (star-forming or otherwise) are 0.94 and 1.12 cMpc, respectively; the highest quartiles ($\ge$ 75th percentile) of $\rperi$ and $\rnow$ are 3.4 and 3.6 cMpc, respectively.

The quenched fraction measured using galaxies that fall in the lowest quartiles of $\rperi$ and $\rnow$, i.e., the galaxies that are (or have been) close to a more massive neighbor, are systematically and significantly higher than the total quenched fraction. Furthermore, $\rnow$ accounts for most of the difference between these populations and the total quenched fraction.  This suggests that interactions with massive neighbors played a role in quenching these systems' star formation, and that current environment alone is a fairly accurate proxy for this effect. In addition, we find that the quenched fraction decreases with increasing stellar mass, suggesting that more massive dwarf galaxies are better able to resist being quenched regardless of $\rperi$. 

\begin{center}
    \begin{figure}[!t]
        \centering
    	\includegraphics[scale=0.34]{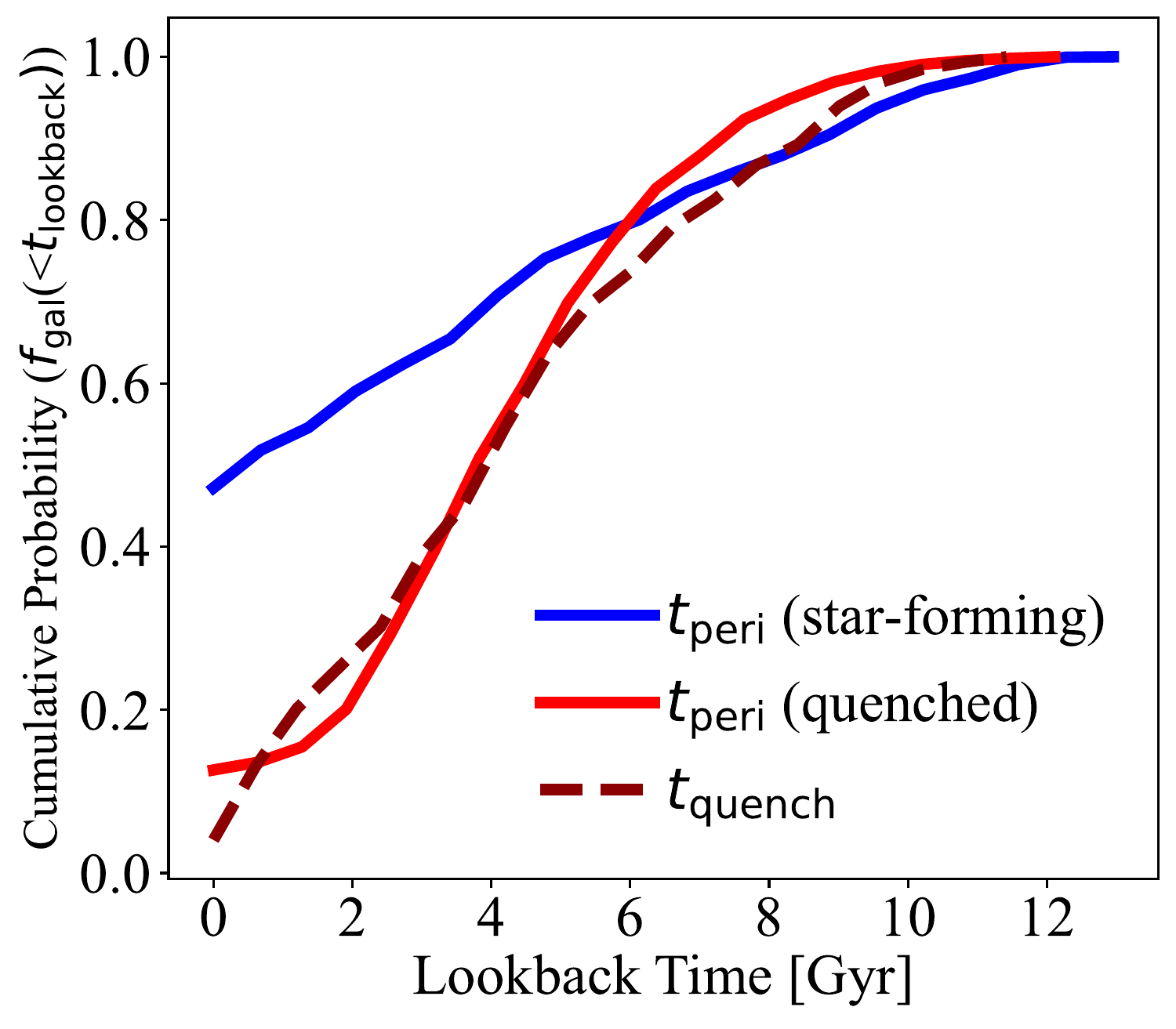}
    	\caption{Cumulative distribution function of the quenching time ($\tquench$, maroon, dashed), time of pericentric passage for star-forming (blue) and quenched (red) populations. $\tquench$ is the lookback time at which the specific star formation rate dropped below the quenching criteria and did not rise again up to and including $z=0$. $\tperi$ is the lookback time at which $\rperi$ occurred. We note that all star-forming systems exist along the $\tquench$ = 0 line and was excluded. Systems with $\tquench>\tperi$ are quenched before reaching $\rperi$, and all systems with $\tquench<\tperi$ are quenched after reaching $\rperi$. For quenched systems, $\tperi$ precedes $\tquench$, suggesting some delay between pericentric passage and quenching. This is consistent with our finding that current environment largely explains offsets from the total quenched fraction.}
        \label{fig:tquenchtperi}
    \end{figure}
\end{center}

\begin{center}
    \begin{figure*}[!t]
        \centering
    	\includegraphics[width=\textwidth]{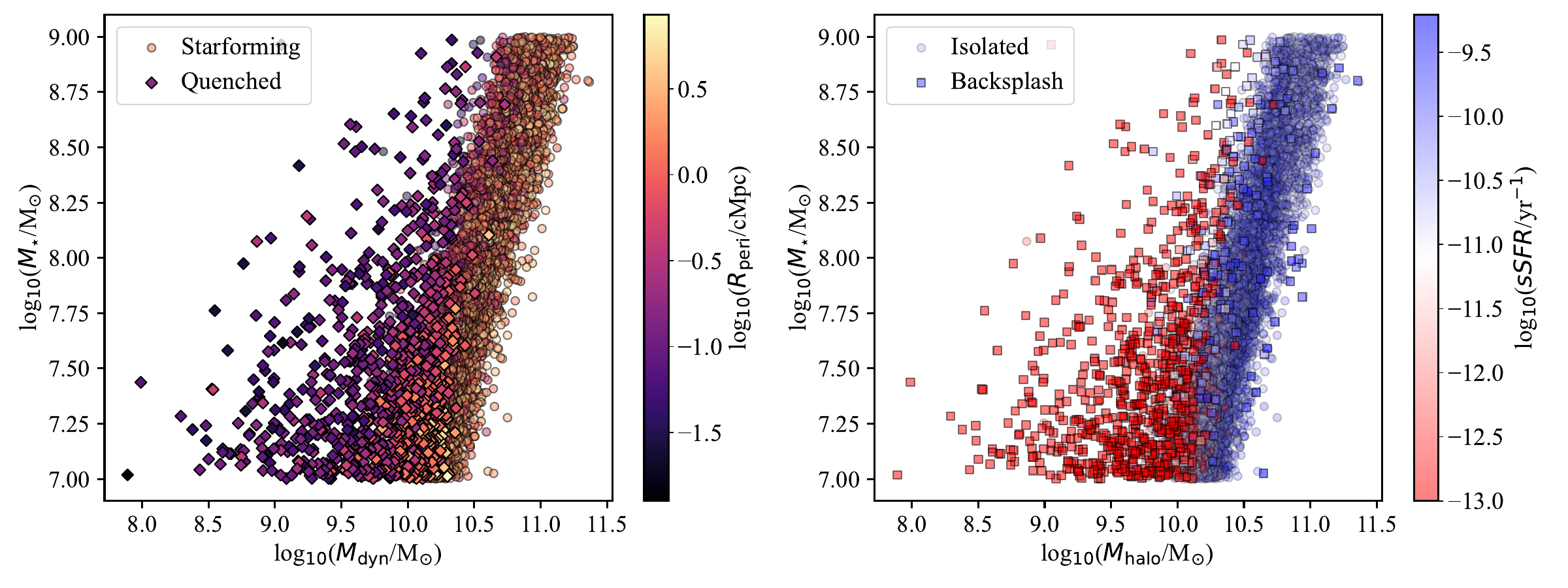}
    	\caption{\emph{Left}: Stellar--halo mass relation for isolated dwarfs, colored by  $R_{\mathrm{peri}}$ and separated into star-forming (circles) and quenched (diamonds) populations. At a fixed stellar mass, galaxies with lower $R_{\mathrm{peri}}$ tend to have lower halo masses, implying that they have been tidally stripped. \emph{Right}: Same as the left panel, but colored by specific star formation rate, split into primaries (circles; i.e., non-backsplash objects) and backsplash systems (squares; i.e., galaxies that have been satellites in the past but are not at $z=0$). Galaxies with the lowest values of $R_{\mathrm{peri}}$ are predominantly quenched backsplash galaxies.}
        \label{fig:shmrdouble}
    \end{figure*}
\end{center}

On the other hand, the quenched fraction measured using galaxies that fall in the highest quartiles of $\rperi$ or $\rnow$, i.e., the galaxies that are far away from a more massive neighbor, is low ($\lesssim 5\%$) but nonzero in the lowest stellar mass bins. This suggests that a small fraction of low-mass galaxies are quenched by a mechanism that does not directly depend on gravitational interactions with another larger system, and that this mechanism preferentially quenches lower-mass systems. This result agrees with the findings of \cite{Wu26}, who found that internal stellar feedback alone can quench field dwarfs in TNG50. We note that it is possible that not all of these galaxies will be permanently quenched; for example, recent work finds that even low-mass dwarf galaxies can ``reignite'' and resume star formation after quenching by reionization or environmental processes~\citep{Wright19,Rey20,Moreno26}. 

The backsplash fraction slowly rises with decreasing stellar mass and surpasses the total quenched fraction in the highest stellar mass bins. This is consistent with larger systems being better able to retain their gas during the infall and ejection process, while smaller-mass galaxies are increasingly likely to be quenched by this kind of interaction. Note that the backsplash fraction does not surpass the quenched fraction in the lowest stellar mass bins, which implies that there are other galaxies at these masses that are quenched by non-backsplash processes. Similar to the trend for backsplash galaxies, these non-backsplash systems are increasingly sensitive to quenching as their mass decreases. 

\subsection{Quenching Times}
\label{quenchingtimes}

In Figure~\ref{fig:tquenchtperi} we compare the cumulative distributions of the lookback times when the quenching criterion defined in Eq.~\ref{eq:sfr_threshold} was most recently satisfied ($\tquench$, dashed line) and when $\rperi$ was achieved ($\tperi$, solid lines) for our sample of star-forming (blue) and quenched (red) isolated dwarf galaxies. The majority of the isolated quenched dwarfs identified at the present-day have $\tquench$ and $\tperi$ values within the last 4 Gyr. Furthermore, many of these quenched dwarfs are also backsplash galaxies, suggesting that recent environmental interactions are responsible for the accumulation of isolated quenched dwarfs in TNG50 at $z=0$. While the shapes of the $\tquench$ and $\tperi$ distributions for quenched dwarfs are similar, we find that $\tquench$ typically occurs at later times compared to $\tperi$. This is likely due to a time delay between a galaxy undergoing pericentric passage and the resultant reduction in star formation, which we find corresponds to a delay of order $\sim 1$ Gyr. This offset is consistent with previous findings of a $\sim 2$ Gyr quenching time for dwarfs beginning at first infall \citep{Wetzel15, Akins21, Greene23, Bhattacharyya25}. 

Figure~\ref{fig:tquenchtperi} also shows that nearly half of the star-forming dwarfs experience $\tperi$ at the present-day, indicating that these systems are either sufficiently far from a more massive neighbor or on their first infall. These dwarfs may either remain isolated, merge with a host galaxy, survive as intact satellites, or become future backsplash galaxies. Conversely, only 10$\%$ of the quenched dwarf sample experience $\tperi$ at the present-day. Unlike the star-forming population, these galaxies were able to quench without having a close interaction with a more massive galaxy. 

Taken together, these findings suggest that the rapid approach to and ejection from the virial radius of a massive halo is the primary mechanism that shut down star formation in the bulk of the isolated quenched dwarfs found in TNG50 at $z=0$. This is corroborated by Figure~\ref{fig:fqquartiles} showing that the backsplash fraction is nearly equal to (and sometimes exceeds) the quenched fraction at all but the lowest stellar masses explored in this study. 

\section{Correlations between Pericentric Distance and Galaxy/Halo Properties}
\label{sec:physical}

\subsection{Stellar Mass--Halo Mass Relation}

Our finding that $\rperi$ strongly correlates with the star formation activity of isolated dwarfs in TNG50 indicates that this metric reliably encodes the environments a galaxy has experienced throughout its evolution. This in itself motivates the comparison of $\rperi$ with other galaxy and halo properties that are likely influenced by environment. We first investigate this in Figure~\ref{fig:shmrdouble} by plotting the stellar--halo mass relation for isolated dwarfs colored by $\rperi$ (left) and by sSFR (right), showing that low $\rperi$ values can accurately predict where quenched systems fall along this relation. 

This result aligns with previous analyses of TNG50; for example, \cite{Bhattacharyya25} found that backsplash galaxies (the systems with the lowest $\rperi$ values in our analysis) occupy lower-mass dark matter halos at fixed stellar mass than systems distant from any massive neighbor today. The backsplash process removes gas and dark matter from galaxies~\citep{Wu26, Benavides25}, explaining why $\rperi$ tracks both sSFR and the stellar mass--halo mass relation. Our result is also qualitatively consistent with analyses of the stellar--halo mass relation across environments in other simulations (e.g., from the MARVEL-ous Dwarfs and D.~C.\ Justice League suites; \citealt{Christensen2024}).

\subsection{$\Delta V_{\mathrm{max}}$ versus Pericentric Distance}

\begin{center}
    \begin{figure}[!t]
        \centering
        \hspace*{-0.2in}
    	\includegraphics[trim={0cm 0cm 0cm 0cm},scale=0.475]{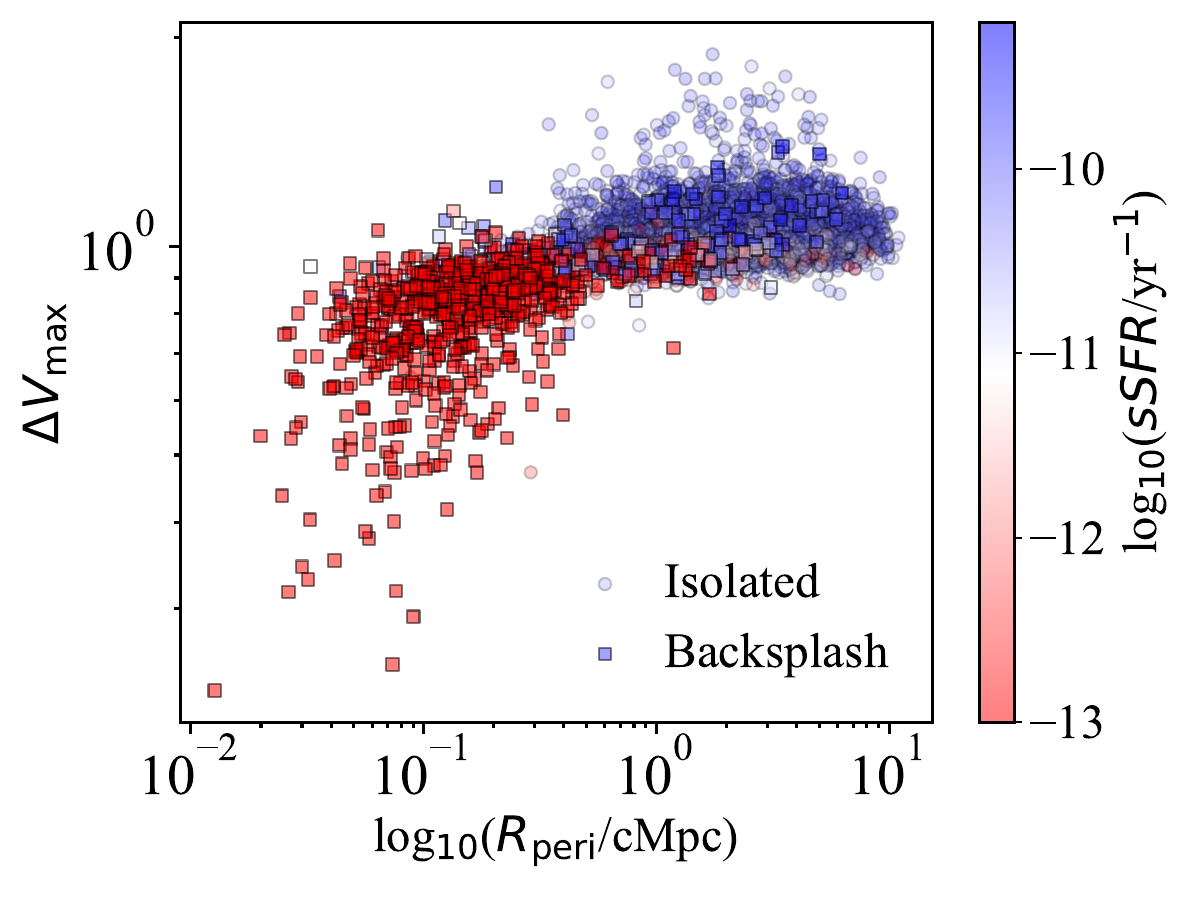}
    	\caption{$\Delta V_{\mathrm{max}}$ vs.\ $\rperi$, colored by specific star formation rate (sSFR). $\Delta V_{\mathrm{max}}$ is the halo assembly parameter--halos with a $\Delta V_{\mathrm{max}} > 1$ have grown over one dynamical timescale, while those with $\Delta V_{\mathrm{max}} < 1$ have been stripped (Eq.~\ref{eq:delta_vmax}; also see \citealt{Wang2024}). The majority of backsplash systems have reduced values of $\Delta V_{\mathrm{max}}$ that are closely related to their star formation rates. Notably, dwarfs in the star-forming backsplash subpopulation have larger values of $\Delta V_{\mathrm{max}}$, suggesting gentler past interactions. Non-backsplash, quenched systems are largely found near $\Delta V_{\mathrm{max}} = 1$, implying a slower quenching process for these systems.}
        \label{fig:vmaxmstar}
    \end{figure}
\end{center}

Figure~\ref{fig:vmaxmstar} shows the relation between $R_{\mathrm{peri}}$ and the halo assembly parameter $\Delta V_{\mathrm{max}}$ defined in Eq.~\ref{eq:delta_vmax}. In the context of the empirical galaxy--halo connection model \textsc{UniverseMachine}, this quantity has been shown to correlate tightly with star formation rate, including in the dwarf galaxy regime~\citep{Wang2021,Wang2024}. Halos with $\Delta V_{\rm{max}} > 1$ have grown over one dynamical timescale, while those with $\Delta V_{\rm{max}} < 1$ have recently been stripped. As indicated by the colorbar in Figure~\ref{fig:vmaxmstar}, TNG50 reproduces this correlation. Furthermore, we find that galaxies with smaller values of $R_{\mathrm{peri}}$ (which, as we have shown, are predominantly backsplash systems) have systematically lower values of $\Delta V_{\rm{max}}$. This follows because backsplash galaxies undergo significant tidal stripping, lowering their present-day values of $V_{\mathrm{max}}$.

These findings provide a physical interpretation for the rise in the quenched fraction at low masses predicted by \cite{Wang2024} by calibrating \textsc{UniverseMachine} to isolated and satellite dwarf galaxy populations. In particular, the rise in the quenched fraction at low masses is tightly linked to orbital history in our framework. 

\section{Discussion}
\label{sec:discussion}
Based on our comparisons of $\rperi$ to environmental and internal properties of isolated dwarf galaxies, we now discuss these results in the context of previous analyses of TNG50 (\S\ref{sec:tng50}) and survey data (\S\ref{sec:observations}); we then show how the dwarf galaxy stellar mass--size relation can be used as an additional proxy for $\rperi$ (\S\ref{sec:size}).

\subsection{Comparison to Other TNG50 Results}
\label{sec:tng50}

We first compare our results to existing works that characterize the isolated quenched dwarfs in TNG50. These comparisons provide context for the correlations between $\rperi$, SFR, and halo properties studied here. 

\cite{Benavides25}, using an identical dwarf stellar mass range and quenching criteria as our work, found that $\approx15\%$ of all isolated dwarfs in the stellar mass range we consider are quenched, with the majority of these quenched systems being backsplash galaxies. They found that the non-backsplash systems were likely quenched by cosmic web stripping, which does not require direct interaction with a massive neighbor and most strongly affects lower-mass systems. Our result in Figure~\ref{fig:fqquartiles} supports this picture---the backsplash fraction is lower than the quenched fraction in the lowest stellar mass bin we consider, and the quenched fraction for the upper $\rperi$ quartile is slightly above zero in this bin. Together, these results imply that another quenching mechanism, which does not involve close approaches to a massive galaxy, subtly raises the quenched fraction of low-mass galaxies. This population is also apparent in Figure~\ref{fig:rnow_rperi} as quenched systems (red circles) close to the $\rnow = \rperi$ relation. These galaxies are consistent with being quenched by a mixture of cosmic web-stripping and internal mechanisms, which we do not attempt to differentiate in this work, unlike \cite{Benavides25} and \cite{Wu26}.

\cite{Bhattacharyya25} also analyzed the stellar--halo mass relation in TNG50 as a function of environment and found that backsplash systems tend to have lower halo masses and higher values of $\Delta$sSFR (which suggests quenching). Our Figure~\ref{fig:shmrdouble} plots these quantities but is colored by $\rperi$. We see, as before, that a low $\rperi$ is associated with backsplash systems; furthermore, $\rperi$ neatly tracks sSFR. This relationship between $\rperi$ and physical parameters like stellar and halo mass will aid in identifying backsplash systems in future surveys. In addition, \cite{Bhattacharyya25} analyzed the relation between dwarf galaxy properties and their current distance to a massive neighbor; their parameterization is identical to our $\rnow$, except that their minimum stellar mass for massive neighbors is $10^{9.5}~\msun$ (versus our threshold of $> 10^{10}~\msun$); they found that this distance measure correlates strongly with dwarf quenching, as suggested by our results. In addition, \cite{Benavides25} found that half of all isolated quenched dwarfs down to $M_*=10^7~\msun$ have a $z=0$ distance from a massive host that is less than 1.25 Mpc. Finally, using the empirical UniverseMachine galaxy--halo connection model calibrated to SAGA satellite population data, \cite{Wang2024} predicted that the quenched fraction of all (satellite and field) dwarfs at $M_*\sim 10^{6.5}~\msun$ is $\approx0.3$. This prediction is in line with our own findings for our lowest-considered stellar mass bin. 

\cite{Shread26} compared dwarf galaxy quenched fractions per stellar mass across different environments in TNG50. Their sample included dwarf galaxies between $1 < D/\mathrm{Mpc} < 25$ (3D distance) of Local Group analogues (using the criteria from \citealt{Pillepich24}) at $z=0$. Group and field galaxies were sorted by a group-finding algorithm created by \cite{Yang05}. According to this algorithm, isolated/field dwarfs are those that are not in a group with any other galaxy, as well as those in a multi-member group that have a distance $> 2~R_{200}$ from the central galaxy and a relative velocity $>3\sigma_v$ with respect to the central, where $R_{200}$ and $\sigma_v$ are the radius velocity dispersion of the main halo, respectively. If a galaxy does not meet either of these criteria, it is assigned to the group that minimizes its relative distance and line-of-sight velocity. Group/non-isolated dwarfs are those that exist in a group with size greater than one. Applying these criteria and calculating the quenched fractions for the isolated dwarfs in their sample yields a quenched fraction that increases with decreasing stellar mass, starting from a minimum near 0 at $~10^8~\msun$, rising to a maximum value of $\sim 0.4$ (for their $1<D/Mpc<25$ sample) in the lowest mass bin we consider. This is supported by our own results for central galaxies, which also exhibit an increasing quenched fraction with decreasing stellar mass, in contrast to existing observational results.

\subsection{Comparison to Survey Data}
\label{sec:observations}

Isolated quenched dwarfs have been identified both individually and as populations in extragalactic surveys. In this section, we will compare the quenched fractions found in existing survey data to our own results. We also discuss the feasibility of identifying the orbital properties of isolated dwarf galaxies (and thus inferring their $\rnow$ and $\rperi$ values), which would inform their histories and potential past interactions with more massive galaxies.  

Considering extremely isolated dwarf galaxies, \cite{Geha12} found that the quenched fraction of dwarfs in SDSS data approaches zero for stellar masses below $\mstar \sim 10^9~\msun$ and systems that are separated from a massive neighbor by more than $4$ virial radii. Conversely, using a mass-complete sample of dwarf galaxies ($\mstar=10^{7-9}~\msun$) in the COSMOS field---which is deeper than both SDSS and SAGAbg---\cite{Lazar26} found that the quenched fraction \emph{increases} below a minimum at $\sim 10^{8.5}~\msun$, reaching $\approx 0.5$ at a minimum stellar mass of $10^7~\msun$. These findings are qualitatively consistent with our TNG50 results and suggest that the relatively shallow data from SDSS and SAGA potentially \emph{underrepresent} the isolated quenched dwarf population. Upcoming data from wide and deep surveys (e.g., LSST and Roman) will be required to confirm this picture. 

Recent observations have pushed the boundaries of isolated quenched dwarf galaxy detectability at lower stellar masses. At the smallest end ($M_{\star} \approx 10^{5-6} M_{\odot}$, some of the furthest observed isolated quenched dwarf galaxies with well-constrained distances to massive neighboring systems are at distances of 1.7 Mpc~\citep{Li24} and 2 Mpc~\citep{Sand24} from a nearest massive neighbor or galaxy group, both of which are many times the virial radii of their respective hosts. This roughly agrees with our upper $\rnow$ quartile for isolated quenched dwarfs from TNG50, which has a median $\rnow$ distance of 2.01 Mpc. These kinds of low-mass dwarfs near massive neighbors should have lower values of $\rperi$ and $\rnow$, and thus---according to our work---are likely to be quenched. 

Finally, we note that observations of isolated quenched dwarfs can yield information about their peculiar velocities and any displacement of gas or dark matter that may suggest a past tidal or ram-pressure interaction. Individual quenched dwarf candidates are now being identified at large distances; for example, DGSAT I was recently discovered in the Pisces-Perseus supercluster (heliocentric distance $\sim 78~ \mathrm{Mpc}$) and may be a backsplash system \citep{MartinezDelgado16}. \cite{Hai26} have also found evidence of isolated dwarfs being quenched by previous interactions with a host due to its position in size-mass space as well as its
low surface brightness which suggests past environmental processing. In general, backsplash systems can be identified based on their peculiar velocities relative to the host group or cluster, and/or by measuring their Sersic profiles and axis ratios as a proxy for previous tidal interactions. Galaxies that are unlikely to have been quenched by backsplashing from a massive neighbor have been tentatively identified as well \citep{Bennet25, Bidaran25, Fielder25, Luber25, McQuinn24, Sand22, Polzin21}. Quantifying the tidal distortion of isolated quenched dwarf candidates, measuring gas distributions, and measuring the peculiar velocity relative to potential host will be necessary to create a framework for determining the history of a quenched dwarf based on observational parameters.

Some work has already been done to infer the orbital histories of isolated Local Group dwarfs. For example, \cite{Bennet25} used measurements of proper motion and orbital modeling from Gaia and HST (method developed in \citealt{Bennet24}) to infer the orbital histories of six nearby isolated dwarfs. They identified one possible backsplash galaxy and five galaxies that are unlikely to have had a pericentric passage in the last 6 Gyr with $>90\%$ certainty. Combining this type of galaxy-by-galaxy analysis with statistical simulation predictions like those presented here is a promising avenue for interpreting upcoming survey data.

\subsection{The Size--Stellar Mass Relation}
\label{sec:size}

\begin{center}
    \begin{figure}[!t]
        \centering
        \hspace*{-0.24in}
    	\includegraphics[trim={0cm 0cm 0cm 0cm},scale=0.375]{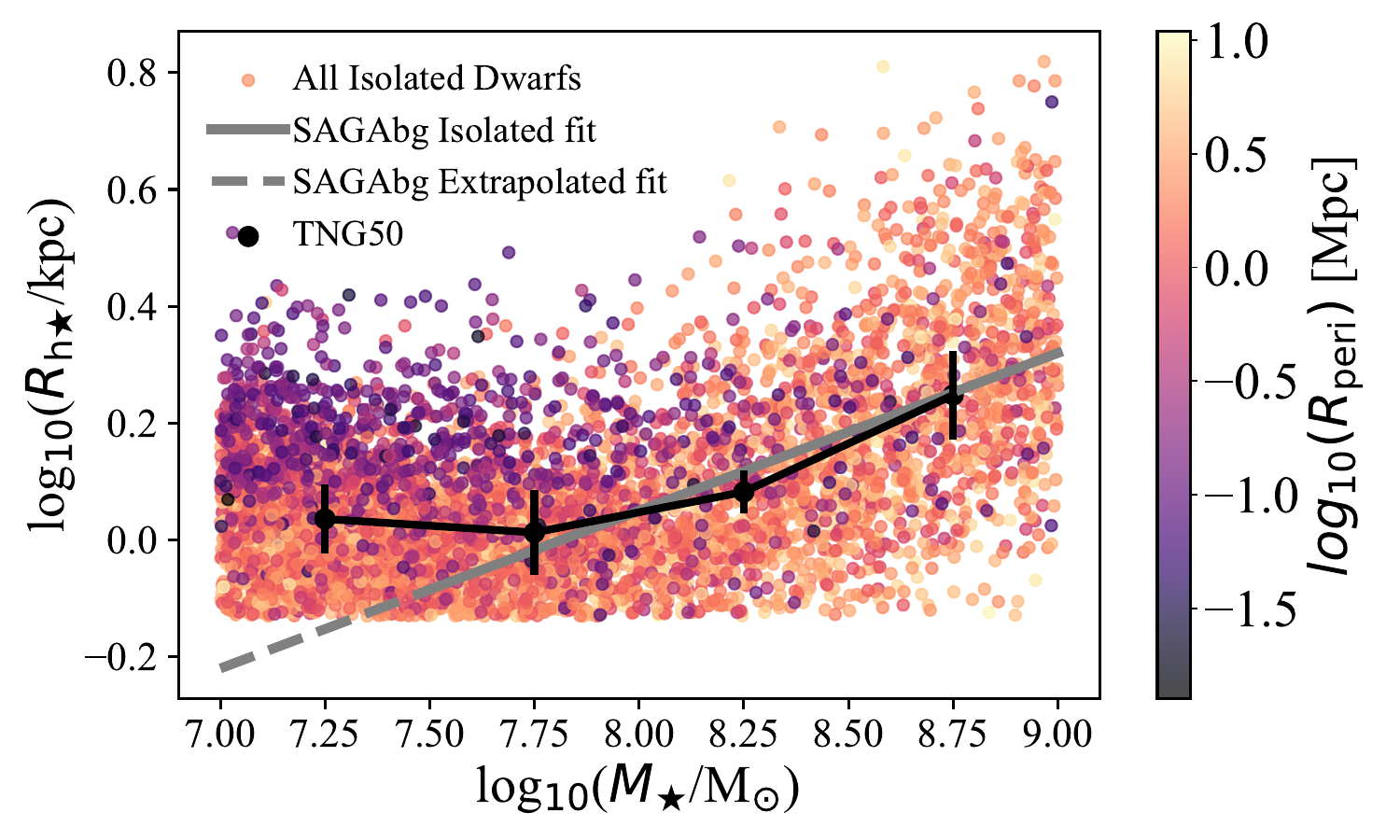}
    	\caption{Size-mass relation of isolated dwarfs in TNG50. The solid grey line is the best fit relation found by \cite{Asali25} for their SAGAbg sample. The dashed grey extension is extrapolates this fit to lower stellar masses. Median size per 0.5 dex mass bin for our TNG50 is in black, with error bars denoting 16th-84th percentiles. The size--mass relation for TNG50 isolated dwarfs rises at the lowest stellar masses we consider, where the  quenched fraction is highest; these systems generally have small $\rperi$ values. The TNG50 relation also diverges from the extrapolated SAGAbg relation in these low stellar mass bins. Note that the TNG50 sample shown here excludes an unphysical ``spur'' of extremely compact galaxies at around $10^8 - 10^9~\msun$ \cite[][see text]{Celiz25}.}\vspace{0.3cm}
        \label{fig:sizemass}
    \end{figure}
\end{center}

Having established that $\rperi$ correlates with quenching and that it is also correlated with halo mass, we briefly study how another key dwarf scaling relation---the size--stellar mass relation---is related to $\rperi$. In particular, Figure~\ref{fig:sizemass} shows the relationship between the size of the galaxies (defined as the radius containing half of the stellar mass) versus total stellar mass. For this analysis, we exclude a subset of very compact, luminous, and mostly star-forming galaxies with $10^8~\msun \lesssim M_*\lesssim 10^9~\msun$. These galaxies were found by \cite{Celiz25} to have more baryonic matter than dark matter at 1 kpc, and sizes r$_{50, \star} \approx 0.74$ ckpc (which is the softening length at $z=0$), which clearly distinguishes them from other galaxies with the same stellar mass. We only exclude this population for our size--mass relation analysis because their star-formation histories are representative of the overall dwarf population.

For the remaining sample, average sizes converge to $\sim 1$ kpc at stellar masses $< 10^8~\msun$. Figure~\ref{fig:sizemass} shows the resulting median size per 0.5 dex mass bin (black) and 16th-84th percentiles (vertical bars) as a function of stellar mass. We find that quenched galaxies have larger sizes than star-forming systems, consistent with their gas and stars being tidally heated. Notably, the dwarf galaxies with the largest sizes that we identify have significantly lower $\rperi$ values at fixed stellar mass. It is likely that galaxies with low $\rperi$ values were affected by tidal stripping or cosmic web stripping; both of these processes can result in larger sizes. This finding is similar to the results of \cite{Mercado25}, who found that galaxies from the FIREbox simulation in low-density environments are less extended than those in high-density environments due to the influence of tidal forces. 

We compare our measurement to the best-fit size--stellar mass relation for the SAGAbg population found by \cite{Asali25} (solid grey), as well as the extrapolation of their fit to the stellar mass bins considered in this work (dashed grey).\footnote{\cite{Asali25} measured galaxy sizes using the effective $r$-band radius $R_{r,\rm{eff}}$, rather than our definition of the half-stellar mass radius $R_{50}$. \cite{xin26} and \cite{vanderWel24} found that $R_{r,\rm{eff}}$ and $R_{50}$ are nearly equivalent at $z\approx0$ using JWST/NIRCam data, so we do not expect this difference to significantly impact the comparison in Figure~\ref{fig:sizemass}.} We find that the SAGAbg data under-represents larger, quenched galaxies compared to our results, especially at the lowest stellar masses considered in this work. In particular, our median size relation shows an upturn between stellar masses of $10^{7-8}~\msun$, suggesting a population of larger, isolated quenched dwarfs that exist below the surface brightness detection limits of SAGAbg data.

However, other existing measurements of the size--mass relation in dwarf galaxies as a function of their current environment contrast our finding that low-$\rperi$ dwarfs have systematically larger sizes. For example, \cite{Mishra23} found an inflection in the size--mass relation below $10^9~\msun$, where the relation breaks to lower sizes, while we find that the size-mass relation inflects toward \emph{larger} sizes at $M_*\approx 10^8~\msun$. Meanwhile, \cite[][ELVES survey]{Carlsten21} and \cite[][GAMA survey]{Lange15} find a log-linear size--mass relation with no inflection down to the lowest stellar masses they measure ($\approx 10^{5.5}~M_{\rm{\odot}}$ and $10^{8}~M_{\rm{\odot}}$, respectively), with \cite{Carlsten21} specifically finding no significant difference in the size--mass relation between red and blue galaxies. \cite{Liao26} found using DESI-LS data that blue (primarily star-forming) galaxies are larger than red (primarily quenched) ones, and that environment at $z \approx 0$ does not play a significant role in determining a galaxy's place on the size-mass relation after controlling for color down to $M_*\approx 10^7~\msun$.

Finally, an analysis of the Fornax Deep Survey from \cite{Watkins23} found a subpopulation of reddened dwarfs in dense environments that are larger due to tidal heating, consistent with our result. Taken together and in the context of our results, these measurements suggest that galaxies' particular orbital histories are not well reflected in their $z\sim0$ survey positions, except for systems in extreme overdensities (e.g., near galaxy clusters). This is plausible given the large scatter between $\rperi$ and $\rnow$ for $\rperi\lesssim 1~\mathrm{Mpc}$ shown in Figure~\ref{fig:rnow_rperi}, but further work that constructs tailored mock samples of TNG50 systems suited to each of these surveys is needed to assess these discrepancies in detail.

Overall, our results suggest that orbital history shapes the size--stellar mass relation of dwarf galaxies. In particular, at fixed stellar mass, galaxies with smaller pericentric distances are systematically more extended, likely due to tidal heating during past interactions. Because these interactions are not uniquely encoded in a galaxy's present-day environment, observational studies based on projected distance or local density alone generally recover only weak correlations between environment and size. This suggests that reconstructing galaxies' orbital histories, or identifying other observational proxies for past pericentric passages, will help interpret the origin of scatter in the dwarf size--stellar mass relation.

\section{Summary $\&$ Conclusion}
\label{sec:conclusion}

Isolated quenched dwarf galaxies are predicted to exist by modern hydrodynamic simulations, and recent observations have uncovered a growing number of these systems at low stellar masses. We anticipate many more to be found in future surveys, and this work was motivated by the need for a framework by which to interpret their observable properties in the context of their orbital histories. A central conclusion of this work is that the past orbital history of an isolated dwarf galaxy, parameterized by its minimum distance to a massive neighbor $\rperi$, is more predictive of its present-day properties than its current environment alone. In particular, we have shown that the likelihood of isolated dwarf quenching is well captured by distance to closest approach to a massive neighbor in the past, $\rperi$, and that $\rperi$ correlates with dwarf galaxy observables including their current distance to a massive neighbor $\rnow$, halo mass and maximum circular velocity, and size.

We summarize our main results as follows: 
\begin{itemize}
\item \textbf{Isolated quenched dwarfs exhibit diverse orbital histories:} Isolated quenched dwarfs in TNG50 at $z=0$ are concentrated at low values of $\rperi$, meaning that these galaxies likely quenched due to previous interactions with a massive neighbor (Figure~\ref{fig:rnow_rperi}). Notwithstanding, isolated dwarfs span the entire distribution of $\rperi$ values, indicating that interactions with a massive galaxy are not required to quench dwarf galaxies. However, only the lowest-mass isolated dwarfs are quenched at large values of $\rperi$ in TNG50 (Figure~\ref{fig:fqquartiles}). 

\item \textbf{The transition between backsplash and non-backsplash quenched systems is set by \boldmath{$\rperi$}:} Galaxies with lower values of $\rperi$ ($\lesssim 0.3$ cMpc) are most likely to be backsplash systems, while those with larger values of $\rperi$ values ($\gtrsim 0.3$ cMpc) are quenched by internal processes or cosmic web stripping (Figure~\ref{fig:rnow_rperi}). 

\item \textbf{Dark matter halo properties correlate with \boldmath{$\rperi$}:} At fixed stellar mass, isolated quenched dwarfs have lower dark matter halo masses compared to their star-forming counterparts (Figure~\ref{fig:shmrdouble}). These systems also generally have lower values of $\rperi$, meaning that it is possible to infer $\rperi$ based on a galaxy's position on the stellar--halo mass relation (and vice versa). Furthermore, low-$\rperi$ dwarfs have experienced more significant recent drops in maximum circular velocity (Figure~\ref{fig:vmaxmstar}), providing a useful foothold for empirical models.

\item \textbf{Scatter in the dwarf galaxy size--mass relation is shaped by \boldmath{$\rperi$}:} For $M_*\lesssim 10^{8}~\msun$, isolated dwarfs in TNG50 with the largest stellar half-mass radii have the lowest values of $\rperi$ (Figure~\ref{fig:sizemass}). These systems are also mostly quenched, suggesting that their extended sizes are associated with previous interactions with a massive neighbor. We also find that low-mass TNG50 dwarfs exhibit, on average, larger half-mass radii relative to expectations from the size--mass relation derived from the SAGAbg sample. This discrepancy may suggest that there is a subpopulation of low-surface-brightness isolated quenched dwarfs that are missing from existing survey data. 

\end{itemize}

Collectively, these results imply that isolated dwarf galaxies retain long-lived signatures of their past environmental interactions, even if they are found in relatively isolated regions today. Future deep imaging surveys, together with dynamical constraints from spectroscopic and proper motion measurements, should therefore constrain the evolutionary pathways of isolated quenched dwarfs by combining multiple observables (including size, halo mass, star-formation activity, and current environment) with physically motivated models of orbital history. We also predict that deep observations from Rubin and Roman should uncover a population of low-surface-brightness, extended, isolated quenched dwarfs that are underrepresented in current dwarf galaxy samples.

Using $\rperi$ as a lens with which to study isolated dwarf galaxies does not reveal every aspect of a galaxy's evolution, especially internal processes that are not captured by dynamical history but may play a role in the star formation of these objects. Nevertheless, our results suggest that the majority of isolated quenched dwarfs are likely affected by either backsplash interactions with another massive galaxy or cosmic web stripping.

The discussion above treats different quenching mechanisms in isolation; however, internal (feedback) and external (tidal) processes may act together. For example, tidal interactions can trigger bursts of star formation that result in supernova feedback, which subsequently quench the galaxy from the inside out. In our framework, it may be possible to predict such events by modeling them probabilistically based on orbital history and thus more accurately predict quenching during gravitational interactions. This type of hybrid model would narrow the gap between external dynamics and internal galactic processes and represents an interesting area for future work. 

Lastly, beyond the study of very isolated galaxies, $\rperi$ may be useful in predicting the evolution of dwarfs in groups or clusters, including satellite galaxies. Dwarf galaxies dominate groups by number, and the distance of closest approach for dwarf galaxies in a group environment (satellite or otherwise) may provide an efficient way to predict dwarf galaxy star formation properties across a range of environments.

\section*{Acknowledgments}
\label{sec:acknowledgements}
We are grateful to Francisco Mercado, Evangela Shread, and Zewei Wu for comments on the manuscript, and we thank José Benavides and Annika Peter for helpful discussions related to this work. DCB is supported by an NSF Astronomy and Astrophysics Postdoctoral Fellowship under award AST-2303800. DCB is also supported by the Future Faculty in the Physical Sciences Postdoctoral Fellowship.
    
\bibliographystyle{apsrev4-1}

\bibliography{oja_template}

\end{document}